\documentclass[aps,showpacs,twocolumn,superscriptaddress,prb,reprint]{revtex4-1}

\usepackage{graphicx}
\usepackage{amsmath,amssymb,amsfonts}
\usepackage{textcomp}
\usepackage{hyperref}
\usepackage{gensymb}
\usepackage{verbatim}
\hypersetup{breaklinks=true,colorlinks=true,urlcolor=black}
\usepackage{ulem}

\begin{document}

\title{Non-monotonic pressure evolution of the upper critical field in superconducting FeSe}

\author{Udhara~S.~Kaluarachchi}
\affiliation{Ames Laboratory US DOE, Iowa State University, Ames, Iowa 50011, USA}
\affiliation{Department of Physics and Astronomy, Iowa State University, Ames, Iowa 50011, USA}
\author{Valentin~Taufour}
\affiliation{Ames Laboratory US DOE, Iowa State University, Ames, Iowa 50011, USA}

\author{Anna~E.~B\"ohmer }
\affiliation{Ames Laboratory US DOE, Iowa State University, Ames, Iowa 50011, USA}
%\affiliation{Institut f\"ur Festk\"orperphysik, Karlsruhe Institute of Technology, 76021 Karlsruhe, Germany}

\author{Makariy~A.~Tanatar}
\affiliation{Ames Laboratory US DOE, Iowa State University, Ames, Iowa 50011, USA}
\affiliation{Department of Physics and Astronomy, Iowa State University, Ames, Iowa 50011, USA}

\author{Sergey~L.~Bud'ko}
\affiliation{Ames Laboratory US DOE, Iowa State University, Ames, Iowa 50011, USA}
\affiliation{Department of Physics and Astronomy, Iowa State University, Ames, Iowa 50011, USA}

\author{Vladimir~G.~Kogan}
\affiliation{Ames Laboratory US DOE, Iowa State University, Ames, Iowa 50011, USA}

\author{Ruslan~Prozorov}
\affiliation{Ames Laboratory US DOE, Iowa State University, Ames, Iowa 50011, USA}
\affiliation{Department of Physics and Astronomy, Iowa State University, Ames, Iowa 50011, USA}

\author{Paul~C.~Canfield}
\affiliation{Ames Laboratory US DOE, Iowa State University, Ames, Iowa 50011, USA}
\affiliation{Department of Physics and Astronomy, Iowa State University, Ames, Iowa 50011, USA}

\begin{abstract}

		The pressure dependence of the upper critical field, $H_\textrm{c2,c}$, of single crystalline FeSe was studied using measurements of the inter-plane resistivity, $\rho_{\textrm{c}}$ in magnetic fields parallel to tetragonal $c$-axis. $H_\textrm{c2,c}(T)$ curves obtained under hydrostatic pressures up to $1.56$\,GPa, the range over which the superconducting transition temperature, $T_\textrm{c}$, of FeSe exhibits a non-monotonic dependence with local maximum at $p_1$\,$\approx$\,0.8\,GPa and local minimum at $p_2\,$$\approx$\,1.2\,GPa. The slope of the upper critical field at $T_\textrm{c}$,  $\left(\textrm{d}H_\text{c2,c}/\textrm{d}T\right)_{T_\textrm{c}}$, also exhibits a non-monotonic pressure dependence with distinct changes at $p_1$ and $p_2$. For $p<p_1$ the slope can be described within multi-band orbital model. For both $p_1<p <p_2$ and $p>p_2$ the slope is in good {\it semi-quantitative} agreement with a single band, orbital Helfand-Werthamer theory with Fermi velocities determined from Shubnikov-de Haas measurements. This finding indicates that Fermi surface changes are responsible for the local minimum of $T_\textrm{c}(p)$ at $p_2$\,$\approx$\,1.2\,GPa.

\end{abstract}
\maketitle

\section{Introduction}
		Hydrostatic pressure is a widely used tool to study materials without changing their stoichiometry. Pressure is a particularly useful, non-thermal tuning parameter for quantum critical materials, in which suppression of antiferromagnetic ordering temperature $T_\textrm{N}$ to zero at a quantum critical point\,\cite{Mathur1998Nature} leads to strong deviations of electronic properties from standard Fermi-liquid theory and superconductivity. 
		Iron-based superconductors provide one of the most clear examples of quantum critical systems\,\cite{Nakai2010PRL}, with $T$-linear resistivity \,\cite{Kasahara2010PRB,Leyraud2009PRB} and maximum $T_\textrm{c}$ at optimal doping found at the edge of magnetic ordering in $x$ or pressure $p$. However, notable deviations from this simple picture are found in K- and Na- hole- doped BaFe$_2$As$_2$ based compositions (BaK122 and BaNa122 in the following). Here suppression of $T_\textrm{N}(x)$ to zero happens at significantly lower $x$ than maximum $T_\textrm{c}$ is achieved\,\cite{Rotter2008PRL,Avci2011PRB,Avci2012PRB,Tanatar2014PRB}, and, moreover, $T_\textrm{c}$ reveals a non-monotonic composition dependence near $x$\,$ \approx$\,0.25\,\cite{Bohmer2014PRL,Allred2015PRB}, and pressure dependence for close compositions\,\cite{Hassinger2012PRB,Budko2013PRB_b,Budko2014PRB}, reminiscent of the 1/8 anomaly in the underdoped cuprates\,\cite{Moodenbaugh1988PRB}. This $x$\,=\,0.25 anomaly in $T_\textrm{c}(x,p)$ was related with the emergence of competing magnetic phase\,\cite{Allred2015PRB}.

		FeSe is structurally the simplest iron based superconductor\,\cite{Paglione2010Nat,Johnston2010,Stewart2011RMP}, but it has one of the more complex pressure-temperature phase diagrams. At ambient pressure FeSe undergoes electronic nematic tetragonal to orthorhombic structural transition with $T_\textrm{s}$\,$\approx$\,90\,K, which is not accompanied by long range magnetic ordering\,\cite{McQueen2009PRL}, and becomes superconducting with $T_\textrm{c}$\,$\approx$\,8.5\,K\,\cite{Hsu2008}. The application of quasi-hydrostatic pressure leads to a four-fold increase of $T_{\textrm{c}}$ up to $37$\,K\,\cite{Miyoshi2014}, and the rise of $T_\textrm{c}$ continues well beyond the point where $T_\textrm{s}(p)\to0$ at $p$\,$\sim$\,2\,GPa. Even much higher $T_{\textrm{c}}$ values up to $\sim$\,100\,K are claimed in single layer films of FeSe\,\cite{Ge2015Nat}. Interestingly, the increase of $T_\textrm{c}$ with pressure in bulk FeSe is not monotonic: $T_\textrm{c}(p)$ shows a local maximum at $p_1\approx0.8$\,GPa, a local minimum at $p_2\approx1.2$\,GPa, before rising monotonically with further pressure increase up to $\approx$\,8\,GPa. 
		The origin of this non-monotonic pressure evolution of $T_\textrm{c}$ in FeSe is a subject of intense studies. The local maximum of $T_\textrm{c}$ was related with emergence a competing phase\,\cite{Terashima2015}, presumably of magnetic origin as observed in $\mu$SR and NMR studies\,\cite{Bendele2010PRL,Bendele2012PRB,Imai2009PRL}. The minimum at $p_2$ (strongly resembling pressure anomaly in the underdoped BaK122\,\cite{Hassinger2012PRB}) can be merely a restoration of a rise of $T_\textrm{c}$ with pressure after decrease at $p_1$, or represent a modification of the superconducting gap structure. 

        A non-monotonic variation of $T_\textrm{c}$ under pressure was also observed in KFe$_2$As$_2$\,\cite{Tafti2013Nat,Terashima2014PRB_b,Taufour2014PRB}. Taufour {\it et al.}\,\cite{ Taufour2014PRB} used measurements of the upper critical field to gain insight into the origin of this anomaly and suggested modification of the superconducting gap structure. Considering the complex evolution of the superconducting transition temperature with pressure in FeSe, measurements of \(H_\textrm{c2}\) can  shed light on the pressure  evolution of the superconducting state of this material.

        In this article we report study of the pressure evolution of the orbital upper critical field $H_\textrm{c2,c}$ as a probe of superconductivity in FeSe. We use inter-plane resistivity measurements in longitudinal configuration with parallel current and magnetic field, $H \parallel j \parallel c$ to minimize the contribution of flux flow phenomena and obtain sharp superconducting transitions. We find a  semi-quantitative agreement for the experimental slope,  $\left(\textrm{d}H_\textrm{c2,c}/\textrm{d}T\right)_{T_\textrm{c}}$,  evaluated at the superconducting transition temperature $T_{\textrm{c}}$ and single\,-\,band Helfand-Werthamer (HW)\,\cite{Helfand1966} calculations using Fermi velocities determined from recent Shubnikov-de Haas (SdH) oscillations\,\cite{Terashima2015arXiv} for $p>p_1$, and  even rough agreement in the multi-band case for $p<p_1$. Three pressure ranges with the characteristic behavior of $T_\textrm{c}(p)$ can be linked with the changes of $H_\textrm{c2,c}$ and of the Fermi surface.

\section{Experimental Methods}
        Single crystals of FeSe were grown using a modified chemical vapor transport technique\,\cite{Boehmer2013PRB}. The $c$-axis resistivity of FeSe was measured using a two probe technique\,\cite{Tanatar2009PRB,Tanatar2009PRB_b} relying on negligible contact resistance. Two Ag wires (50\,$\mu$m diameter) were attached to the samples by soldering with  In-Ag alloy, giving contacts with resistance less than 10$\mu\Omega$. Four-probe measurements were used down to the sample contacts, so that the measured resistance represents sum of series connected sample and contact resistances, $R_{sample}+R_{contact}$. Because $R_{contact} \ll R_{sample}$, the contact resistance gives minor correction, of order of 1\%, to measured quantity. This can be directly seen from negligible measured resistance at temperatures below superconducting transition of FeSe, see Fig.1 below. Measurements were performed in a {\it Quantum Design} PPMS, on cooling and warming at a rate 0.25 K/min. We used a Be-Cu/Ni-Cr-Al hybrid piston-cylinder cell, similar to the one described in Ref.\,\onlinecite{Budko1984}. Pressure values at low temperatures were inferred from $T_\textrm{c}(p)$ of lead\,\cite{Bireckoven1988}. Good hydrostatic pressure conditions, as seen by sharp superconducting transitions of both sample and Pb resistive manometer, were achieved by using a pressure medium of 4:6 mixture of light-mineral oil\,:\,$n$-pentane\,\cite{Budko1984,Kim2011PRB} that solidifies at room temperature at $\sim$\,3-4\,GPa, well above our maximum pressure. The orientation of the sample in the pressure cell was adjusted so that magnetic field was applied parallel to the $c$\,-axis direction. 
        
\section{Results and Discussion}
        \begin{figure}[!htb]
                \begin{center}
                        \includegraphics[width=8cm]{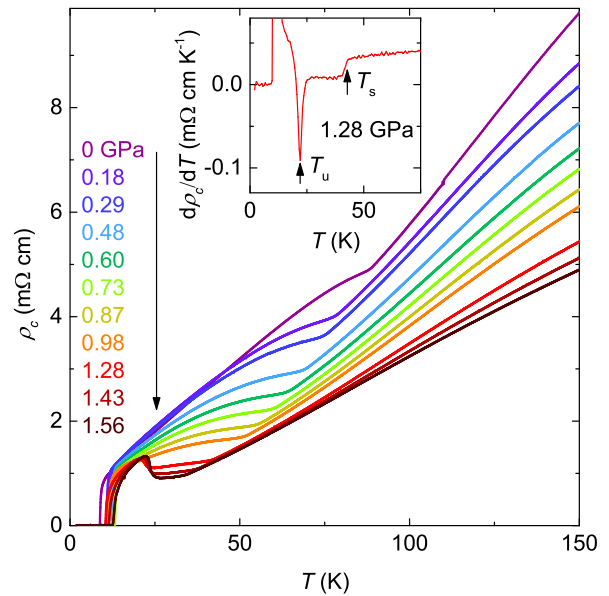}
                \end{center}
                \caption{\label{Rho} (color online) Evolution of the temperature dependence of the $c$-axis resistivity, $\rho_{\textrm{c}}$, with hydrostatic pressure. The inset presents the derivative of $\rho_c(T)$ taken at $1.28$\,GPa, with arrows indicating the temperatures of the structural transition, $T_{\textrm{s}}$, and of the "unknown" transition, $T_{\textrm{u}}$, which is presumably associated  with magnetic ordering.}
        \end{figure}

		Figure\,\ref{Rho} shows the temperature dependence of the $c$-axis resistivity, $\rho_{\textrm{c}}$, taken at various pressures. At ambient pressure the inter-plane resistivity decreases on cooling and shows an anomaly associated with the structural phase transition at $T_{\textrm{s}}$\,$\sim$\,86\,K and a sharp superconducting transition at $T_{\textrm{c}}$\,$\sim$\,9.5\,K. The residual resistivity ratio (RRR) values are found to be 17 for the $c$-axis data shown in this work and 22 for $ab$-plane data shown in Ref.\,\onlinecite{Tanatar2015arXiv}. Similar values of $T_{\textrm{s}}$ and $T_{\textrm{c}}$ have been obtained in previous in-plane resistivity ($\rho_{\textrm{ab}}$) studies\,\cite{Miyoshi2014,Knoner2015PRB,Terashima2015}. Note, however, that the anomaly at $T_{\textrm{s}}$ is much more prominent in $\rho_{\textrm{c}}(T)$. With increasing pressure, the resistivity at $150$\,K monotonically decreases and $T_\textrm{s}$ is also suppressed. A sudden increase of $\rho_{\textrm{c}}$ on cooling below $T_{\textrm{u}}$\,$\sim$\,15\,K, as seen for pressure above $0.87$\,GPa, marks emergence of a new, most likely magnetically ordered\,\cite{Bendele2010PRL,Bendele2012PRB,Terashima2015}, phase. The pressure range of this phase in our experiments is consistent with previous report of Ref.\onlinecite{Terashima2015}, see below. Note that the anomaly at $T_{\textrm{u}}$ is also more prominent in $\rho_c(T)$ than in $\rho_{ab}$. No measurable temperature hysteresis is found for any of the transitions. The values of  $T_{\textrm{s}}$ and $T_{\textrm{u}}$ were obtained from the features in the resistivity derivative, inset of Fig.\,\ref{Rho}. They were tracked as function of applied pressure to determine $p$-$T$ phase diagram of the compound, as shown in Fig.\,\ref{T_P-2nd} below.

        \begin{figure}[!htb]
                \begin{center}
                        \includegraphics[width=8cm]{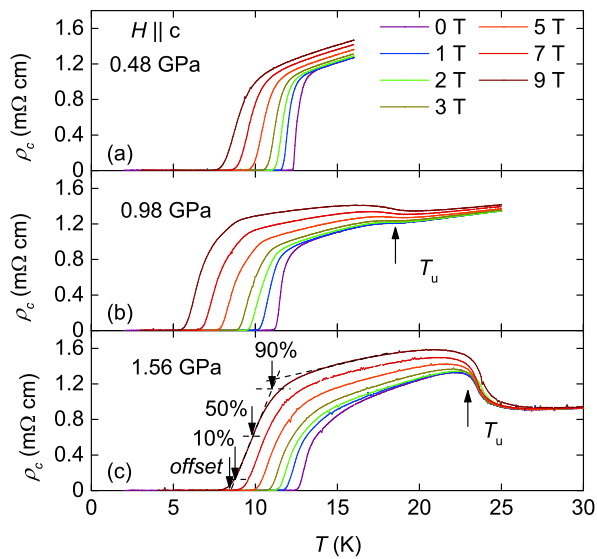}
                \end{center}
                \caption{\label{rho_T-H_p} (color online) Temperature dependence of the inter-plane resistivity, $\rho_{c}(T)$, taken in magnetic fields of 0\,T to 9\,T range in the $H \parallel c$ configuration at pressures of 0.48 GPa (a), 0.98 GPa (b), and 1.56\,GPa (c). The upper critical field $H_\textrm{c2,c}$ is determined from the offset, 10$\%$, 50$\%$, 90$\%$ of $\rho_{\textrm{c}}$ as shown schematically by the arrows in the bottom panel for the 9\,T curve and unless otherwise stated, offset criterion is used in this article. The temperature of the unknown transition, $T_\textrm{u}$,  manifests a very weak dependence on external magnetic fields up to 9\,T.}
        \end{figure}

		Figure\,\ref{rho_T-H_p} shows the evolution of the inter-plane resistivity, $\rho_{\textrm{c}}(T)$, in the vicinity of the superconducting transition with magnetic fields of 0\,T to 9\,T range. The data are shown for representative pressures of 0.48\,GPa (a), 0.98\,GPa (b) and 1.56\,GPa (c). $T_{\textrm{c}}$ was defined using an offset criterion as schematically shown in Fig.\,\ref{rho_T-H_p}(c) for 9\,T curve. At ambient pressure the superconducting transition remains quite sharp for all field values. The transition broadens slightly at higher pressures and magnetic fields. Whereas the $T_{\textrm{c}}$ is suppressed at a similar rate at 0.48\,GPa  and 1.56\,GPa (Figs.\,\ref{rho_T-H_p} (a), (c)), it is suppressed much faster at the intermediate pressure range at 0.98\,GPa (Fig.\,\ref{rho_T-H_p}(b)). We note that $T_{\textrm{u}}$ only shows a weak dependence on applied magnetic field up to 9\,T, see Figs.\,\ref{rho_T-H_p} (b) and (c).

        \begin{figure}[!htb]
                \begin{center}
                        \includegraphics[width=8cm]{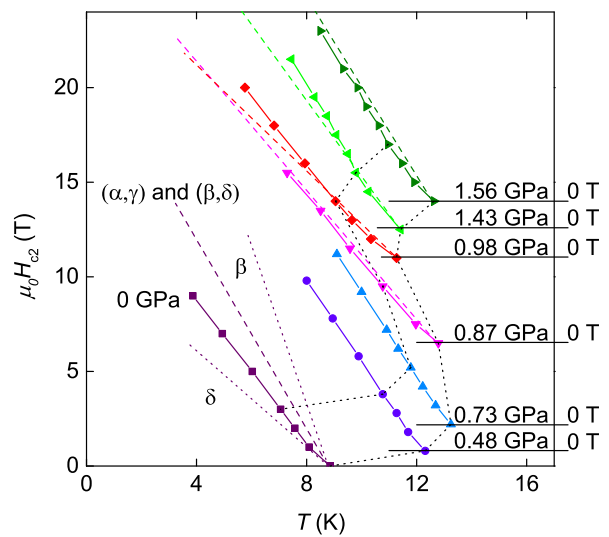}
                \end{center}
                \caption{\label{Hc2_offset} (color online) The temperature dependence of the upper critical field $H_\textrm{c2,c}(T)$ measured in $H \parallel c$ configuration under various pressures. The data are vertically offset to avoid overlapping, with horizontal lines showing $H$\,=\,0. Dashed lines represent upper critical field slopes calculated based on the Fermi velocity, $v_{\textrm{F}}$, from Ref.\onlinecite{Terashima2015arXiv} for a cylindrical Fermi-surface\,\cite{Kogan2012} (see text). At 0\,GPa,  two dotted and dashed  lines represent calculated slopes for the largest  ($\delta$) , smallest ($\beta$) and average $v_{\textrm{F}}$ (($\alpha,\gamma$) and ($\beta,\delta$)). Black dotted lines connect the 0\,T and 3\,T points for various pressures.}
        \end{figure}

		The pressure evolution of the temperature-dependent upper critical field  $\textrm{$\mu $}_{0}H_\textrm{c2,c}$, measured in $H \parallel c$ configuration, is summarized in Fig.\,\ref{Hc2_offset}. The symbols connected by solid lines represent the experimental data, the curves for different pressures are offset to avoid overlapping, and $H=0$ origin for each pressure is shown by solid horizontal lines. We can clearly see that the slope of $H_\textrm{c2,c}$ at $T_{\textrm{c}}$, d$H_\textrm{c2,c}$\,/\,d$T$$\mid_{T_\textrm{c}}$, is fairly constant between 0\,GPa and 0.73\,GPa, decreases markedly when the pressure is increased to 0.87\,GPa and increases again at 1.43\,GPa.

        Generally, the upper critical field is determined by the orbital and Pauli pair breaking effects. Near $T_\textrm{c}$, the Pauli limit is irrelevant and the slope, d$H_\textrm{c2,c}$\,/\,d$T\mid_{T_{\textrm{c}}}$, can be estimated in the clean limit for cylindrical Fermi-surfaces as\,\cite{Kogan2012}:

        \begin{equation}
        -\!\mu_0\frac{\text{d}H_{c2}}{\text{d}T}\bigg|_{T_\textrm{c}}\!\!\!=\!\frac{16\pi k_{\text{B}}^{2}\Phi_{\text{0}}T_\textrm{c}}{7\zeta(3)\hbar^{2}(n_{1}\lambda_{11}\!<\!v_{1}^{2}\!>\!+n_{2}\lambda_{22}\!<\!v_{2}^{2}\!>)} \\
        \label{Eq__Hc2}
        \end{equation}

        \noindent where, $n_{1}=N_{1}/N_{total}$ and $n_{2}=N_{2}/N_{total}$ are the partial densities of states which can be obtained from $N_i\propto m^{*}_ik_{\text{F},i}$. $v_i$ are the Fermi-velocity, $m^{*}_i$ the effective mass and $k_{\text{F},i}$ the Fermi wave vector of the respective band $i=(1,2)$.  $\lambda_{11}$ and the $\lambda_{22}$ are the normalized coupling constant\,\cite{Kogan2012} and $\Phi_{0}$ is the flux quantum.

        To compare the pressure evolution of the upper critical field with changes of the Fermi surface, we calculate the d$H_\textrm{c2,c}$/d$T$$\mid_{T_\textrm{c}}$ with Eq.\,\ref{Eq__Hc2}, using $v_{F}$ values determined in recent quantum oscillation studies by Terashima {\it et al.}\,\cite{Terashima2015arXiv}. The calculated d$H_\textrm{c2,c}$/d$T$$\mid_{T_\textrm{c}}$ represented by the dashed lines in Fig.\,\ref{Hc2_offset} shows semi-quantitative agreement with the experimental slopes. This agreement is good in particularly for pressures above 0.6\,GPa, where only one fundamental frequency is observed in SdH studies. This allows us to select $n_{1}=\lambda_{11}=1$ and $n_{2}=\lambda_{22}=0$ in Eq.\,\ref{Eq__Hc2}. The calculated slopes reproduce very well the experimental data up to the highest pressure 1.56\,GPa of our experiment, including the increase of the slope between 0.98\,GPa and 1.43\,GPa.  The situation is more complicated for pressures below  0.6\,GPa, where four fundamental frequencies, $\alpha$, $\beta$, $\gamma$ and $\delta$ are observed in SdH measurements. As explained in Ref.\,\onlinecite{Terashima2014PRB}, $\alpha$ and $\gamma$ orbits are attributed to the electron like Fermi pockets whereas $\beta$ and $\delta$ to the hole pockets. Following this assignment, we can calculate the average $v_{\textrm{F}}$ and partial densities of states for each Fermi-surface. We consider this system as effective two band case and assume  $\lambda_{11}$\,=\,$\lambda_{22}$ to estimate d$H_\textrm{c2,c}$/d$T$$\mid_{T_\textrm{c}}$ at zero pressure using Eq.\,\ref{Eq__Hc2}. The result is shown by a dashed line in Fig.\,\ref{Hc2_offset}. For reference we also show the estimated slopes using extreme Fermi velocities, the largest - $\delta$ frequency, and the smallest - $\beta$ frequency, as represented by dotted lines. Hence the calculation reproduces the range of value for the slope. We point that slight variation of the coupling constants can improve the match. This comparison suggests that  the pressure evolution of the upper critical field over the whole range can be explained by measured Fermi-velocities.

        This finding contrasts with KFe$_2$As$_2$, where a change in (d$H_\textrm{c2}$\,/\,d$T$)\,/\,$T_\textrm{c}$ as a function of pressure was measured, but could not be attributed to changes of the Fermi velocities\,\cite{Taufour2014PRB}. Whereas this was taken as an indication for a change of the order parameter with pressure in KFe$_2$As$_2$, the case of FeSe seems to be more conventional, as the changes in the upper critical field can be well explained by the observed change of Fermi velocities.

	    \begin{figure}[!p]
			     	\begin{center}
			     		\includegraphics[width=8cm]{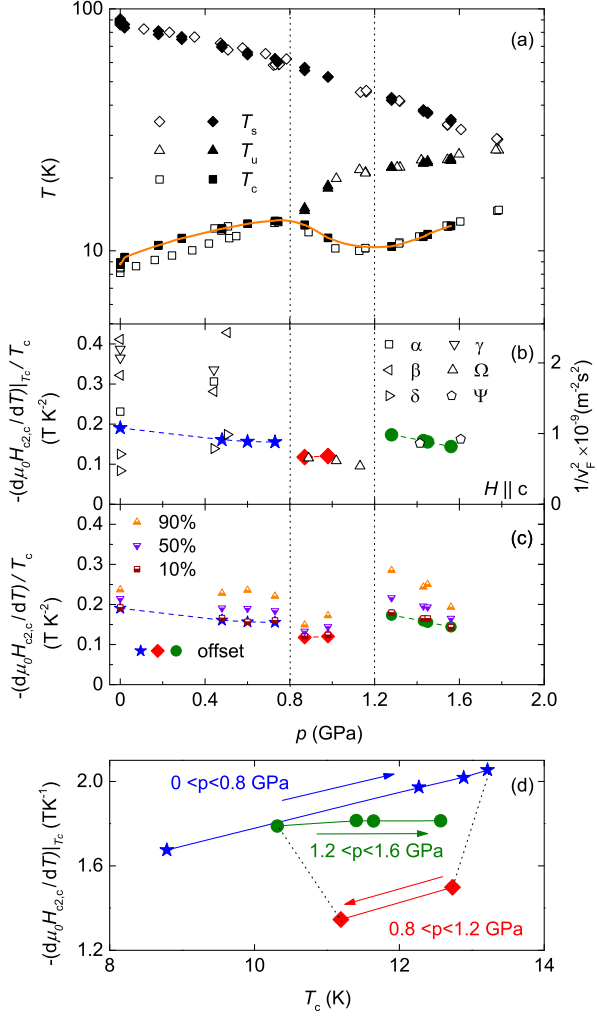}
			     	\end{center}
			     	\caption{\label{T_P-2nd} (color online) (a) Temperature\,-\,pressure phase diagram of FeSe as determined from inter-plane resistivity measurements (full symbols) and from previous in-plane resistivity measurements (open symbols)\,\cite{Knoner2015PRB,Terashima2015,Terashima2015arXiv}. The circles, triangles and squares represent the tetragonal/orthorhombic, "unknown", most likely magnetic and superconducting phase transitions, respectively. The orange solid line is a guide for the eye. Vertical dotted lines show the pressures corresponding to the local maxima and minima of $T_{\textrm{c}}(p)$. (b) Pressure dependence of the normalized slope $(-1/T_\textrm{c})(\text{d}H_{\text{c2},c}/\text{d}T)|_{T_\textrm{c}}$. An abrupt change of slope is observed near 0.8\,GPa and 1.2\,GPa, corresponding to the maximum and a minimum in $T_{\textrm{c}}(p)$ (left axis). For reference we show $v_\textrm{F}^{-2}$ calculated for individual orbits in Shubnikov-deHaas effect measurements of Ref.\,\onlinecite{Terashima2015arXiv} (right axis). There is clear proportionality between the normalized slope and $v_\textrm{F}^{-2}$ as found in our data analysis for $p>p_1$, despite both slope and $v_\textrm{F}$ showing non-monotonic changes at $p_2$. Multiple orbits found for $p<p_1$ clearly illustrate the difficulty of $H_\textrm{c2,c}$ slope calculation in any simple model. (c) Pressure dependence of the normalized slope $(-1/T_\textrm{c})(\text{d}H_{\text{c2},c}/\text{d}T)|_{T_\textrm{c}}$ for four different criteria; offset, 10$\%$, 50$\%$ and 90$\%$ of the resistivity. (d) The slope of the $H_\textrm{c2,c}$ line plotted vs $T_\textrm{c}$ with pressure as a implicit parameter. The plot clearly reveals the non-monotonic dependence with three ranges separated at pressures $p_1$ and $p_2$. Arrows indicate the direction of pressure increase.}
	    \end{figure}

        Figure\,\ref{T_P-2nd} shows the pressure evolution of the slope $(-1/T_\textrm{c})(\text{d}H_{\text{c2},c}/\text{d}T)|_{T_\textrm{c}}$ determined by a linear fit to $H_\textrm{c2,c}$ data in the field interval 0-3\,T, and its relation to the temperature-pressure phase diagram of FeSe. The phase diagram (Fig.\,\ref{T_P-2nd}(a)) as determined from our inter-plane resistivity measurements is in perfect agreement with previous results determined from in-plane resistivity measurements\,\cite{Knoner2015PRB,Terashima2015,Terashima2015arXiv}. The temperature of the nematic transition $T_{\textrm{s}}$ shows a linear decrease with a slope of $\text{d}T_\textrm{s}\text{/d}p\approx\text{-34 K/GPa}$. The superconducting $T_{\textrm{c}}(p)$ has maximum and minimum around $p_1\approx0.8$\,GPa and $p_2\approx1.2$\,GPa respectively. The maximum is located close to the point where anomaly at $T_\textrm{u}$ emerges, likely signaling a magnetic phase transition\,\cite{Bendele2010PRL,Bendele2012PRB,Imai2009PRL}. Competition between superconductivity and magnetic order may be the reason for the suppression of $T_{\textrm{c}}$ between $p_1$ and $p_2$. In contrast, no anomalies which would correlate with either the maximum or the minimum of $T_{\textrm{c}}$ are observed in $T_{\textrm{s}}$.

        The existence of a maximum and a minimum in $T_{\textrm{c}}(p)$ makes us divide the phase diagram into three different regions: $p\lesssim p_1$, (d$T_\textrm{c}$/d$p>0$), $p_1\lesssim p\lesssim p_2$, (d$T_\textrm{c}$/d$p<0$) and $p_2\lesssim p$, (d$T_\textrm{c}$/d$p>0$), as represented by vertical dotted lines in Fig.\,\ref{T_P-2nd}. Interestingly, the pressure evolution of the normalized slope $(-1/T_\textrm{c})(\text{d}H_{\text{c2},c}/\text{d}T)|_{T_\textrm{c}}$ (Fig.\,\ref{T_P-2nd}(b))  shows much more pronounced changes between the three ranges.  In order to demonstrate that our results are not criteria dependent (see Fig.\,\ref{rho_T-H_p}), Fig.\,\ref{T_P-2nd}(c) shows a comparison of the pressure dependence of $(-1/T_\textrm{c})(\text{d}H_{\text{c2},c}/\text{d}T)|_{T_\textrm{c}}$ for  following criteria; offset, 10$\%$, 50$\%$ and 90$\%$ of the resistivity. Due to the curvature at onset of the resistivity data, 90$\%$ criterion shows considerably higher  $(-1/T_\textrm{c})(\text{d}H_{\text{c2},c}/\text{d}T)|_{T_\textrm{c}}$ values than other criteria. However we can clearly see that the overall behavior of pressure dependence of $(-1/T_\textrm{c})(\text{d}H_{\text{c2},c}/\text{d}T)|_{T_\textrm{c}}$ does not depend on the chosen criteria.
        
        The differences between the three pressure ranges are particularly visible when plotting the data as a function of $T_\textrm{c}$ (Fig.\,\ref{T_P-2nd}(d)) with pressure as an implicit hidden parameter. In the low pressure range, $(-1/T_\textrm{c})(\text{d}H_{\text{c2},c}/\text{d}T)|_{T_\textrm{c}}$ remains rather constant, then shows a sudden drop at $p_1\approx0.8$ GPa, where $T_\textrm{c}$ has a maximum, and an increment around $p_2\approx1.2$\,GPa where $T_\textrm{c}$ reaches a local minimum (Fig.\,\ref{T_P-2nd}(b)). Note that in the case of KFe$_2$As$_2$, the abrupt change of the slope $(-1/T_\textrm{c})(\text{d}H_{\text{c2},c}/\text{d}T)|_{T_\textrm{c}}$ also coincides with a minimum in $T_\textrm{c}(p)$\cite{Taufour2014PRB}.

        Simplifying equation\,\ref{Eq__Hc2} for the single-band case, one can relate the initial $H_\textrm{c2}$ slope to the Fermi-velocity, $v_\textrm{F}$, and the effective mass, $m^{*}$, as: 
		$(-1/T_\textrm{c})(\text{d}H_{\text{c2},c}/\text{d}T)|_{T_\textrm{c}}$$\propto v_\text{F}^{-2}$$\propto(m^*)^2$. This dependence allows for direct comparison between the initial slope of $H_\textrm{c2,c}$  and Fermiology. In Fig.\,\ref{T_P-2nd}(b) we plot the normalized slope of $H_\textrm{c2,c}$ (left axis) and $v_\textrm{F}^{-2}$ (right axis). The two quantities show very similar pressure dependence, directly illustrating the validity of our previous analysis. This plot also provides graphical illustration for the difficulty of quantitative comparison of the slope with Fermiology in the multi-band case in particular considering possibility of variation of coupling constants.

        The pressure evolution of $(-1/T_\textrm{c})(\text{d}H_{\text{c2},c}/\text{d}T)|_{T_\textrm{c}}$, hence, indicates a decrease of the effective masses with pressure in the low pressure range, a further sudden decrement around $p_1$, and an increase of the effective mass around $p_2$. This is indeed in good agreement with the quantum oscillation study\,\cite{Terashima2015arXiv}. It is notable, however, that $T_\textrm{c}(p)$ shows only smooth changes around these pressure values.

\section{Conclusions}
        In conclusion, the pressure evolution of the upper critical field of FeSe is in good agreement with the Fermi velocities determined from quantum oscillations. Abrupt changes in the normalized slope of the upper critical field $(-1/T_\textrm{c})(\text{d}H_{\text{c2},c}/\text{d}T)|_{T_\textrm{c}}$ of FeSe provide evidence for changes of the Fermi surface around 0.8 and 1.2\,GPa, which correspond to the local maximum and minimum of $T_{\textrm{c}}$, respectively.  We cannot exclude possible effects on change of order parameter and/or variation of the coupling under pressure may be the reason for the observed change in $T_\text{c}$. However, our study demonstrates that, in contrast to KFe$_2$As$_2$, the non-monotonic pressure evolution of $T_\textrm{c}$ of FeSe   can be   fully accounted for by changes in  the Fermi surface.

	\section*{ACKNOWLEDGMENTS}
        We would like to thank A.\,Kreyssig and T.\,Kong for useful discussions and T.\,Terashima for sharing his quantum oscillation data for the comparison in this study. This work was carried out at the Iowa State University and supported by the Ames Laboratory, U.S. DOE, under Contract No.DE-AC02-07CH11358. V.T. is partially supported by Critical Material Institute, an Energy Innovation Hub funded by U.S. DOE, Office of Energy Efficiency and Renewal Energy, Advanced Manufacturing Office.

%\bibliographystyle{apsrev4-1}
%\bibliography{MyDataBase}

%merlin.mbs apsrev4-1.bst 2010-07-25 4.21a (PWD, AO, DPC) hacked
%Control: key (0)
%Control: author (72) initials jnrlst
%Control: editor formatted (1) identically to author
%Control: production of article title (-1) disabled
%Control: page (0) single
%Control: year (1) truncated
%Control: production of eprint (0) enabled
%

\end{document}